\documentclass[usenatbib]{mn2e}

\usepackage{psfig,morefloats,url}

\footnotesize
\newdimen\digitwidth    
\setbox0=\hbox{\rm0}
\digitwidth=\wd0
\catcode`!=\active
\def!{\kern\digitwidth}
\normalsize

\def\lapp{\ifmmode\stackrel{<}{_{\sim}}\else$\stackrel{<}{_{\sim}}$\fi}
\def\gapp{\ifmmode\stackrel{>}{_{\sim}}\else$\stackrel{>}{_{\sim}}$\fi}

\title[Age constraints in the double pulsar system]
{Age constraints in the double pulsar system J0737--3039}
\author[D.~R. Lorimer et al.]
{D.R.~Lorimer$^1$,\thanks{Email: Duncan.Lorimer@mail.wvu.edu}
P.C.C.~Freire,$^{2}$
I.H.~Stairs,$^{3}$
M.~Kramer,$^{4}$ 
M.A.~McLaughlin,$^{1}$ 
\newauthor
M.~Burgay,$^{5}$ 
S.E.~Thorsett,$^{6}$
R.J.~Dewey,$^{6}$
A.G.~Lyne,$^{4}$ 
R.N.~Manchester,$^{7}$ 
\newauthor
N.~D'Amico,$^{5}$ 
A.~Possenti$^{5}$ and
B.C.~Joshi$^{8}$\\
$^{1}$Department of Physics, West Virginia University, Morgantown, 
WV 26506, USA\\
$^{2}$NAIC, Arecibo Observatory, HC3 Box 53995, Arecibo, PR 00612, USA\\
$^{3}$Department of Physics and Astronomy, University of British Columbia,
6224 Agricultural Road, Vancouver, BC V6T 1Z1, Canada\\
$^{4}$INAF - Osservatorio Astronomica di Cagliari, 09012 Capoterra, Italy\\
$^{5}$University of Manchester,
Jodrell Bank Observatory, Macclesfield, Cheshire, SK11~9DL, UK\\
$^{6}$Department of Astronomy and Astrophysics, University of California,
Santa Cruz, CA 95064, USA\\
$^{7}$ATNF, CSIRO, PO Box 76, Epping, NSW 2121, Australia\\
$^{8}$NCRA, PO Box Bag 3, Ganeshkhind, Pune 411007, India}

\let\leqslant=\leq 
\date{Accepted 2007 May 22. Received 2007 May 7}
\begin{document}

\maketitle
\newcommand{\setthebls}{
}

\setthebls

\begin{abstract} 
We investigate the age constraints that can be placed on the double
pulsar system using models for the spin-down of the first-born 22.7-ms
pulsar A and the 2.77-s pulsar B with characteristic ages of 210 and 50~Myr
respectively. Standard models assuming dipolar
spin-down of both pulsars suggest that the time since the formation of
B is $\sim 50$~Myr, i.e.~close to B's characteristic age.  However,
adopting models which account for the impact of A's relativistic wind
on B's spin-down we find that the formation of B took place either 80
or 180~Myr ago, depending the interaction mechanism. Formation 80~Myr
ago, closer to B's characteristic age,
would result in the contribution from J0737--3039 to the inferred
coalescence rates for double neutron star binaries increasing by
40\%. The 180~Myr age is closer to A's characteristic age and would be
consistent with the most recent estimates of the coalescence rate.
The new age constraints do not significantly impact recent estimates of
the kick velocity, tilt angle between pre and post-supernova
orbital planes or pre-supernova mass of B's progenitor.
\end{abstract}


\section{Introduction}\label{sec:intro}

In addition to its use as a laboratory for studying general relativity
and plasma physics, the double pulsar system J0737--3039
\citep{bdp+03,lbk+04} provides new insights into the evolution of
massive binary systems.  In the standard model for binary pulsar
formation \citep[for a recent review, see][]{van07}, double neutron
star systems are formed from massive binary systems where the
initially more massive (primary) star evolves off the main sequence
and undergoes a supernova explosion to form a neutron star. During the
evolution of the initially less massive (secondary) star, the
first-born neutron star accretes matter and gains angular momentum
spinning it up to short periods and \citep[through poorly understood
processes;][]{smsn89} reducing its magnetic field. If the secondary is
sufficiently massive to explode as a supernova, and the binary system
survives this explosion, the resulting system is a pair of neutron
stars: a short-period recycled pulsar from the primary and a young
``normal'' pulsar from the secondary.

The double pulsar system J0737--3039 where a recycled 22.7-ms pulsar
(hereafter A) is observed in a 2.4-hr orbit around a 2.77-s pulsar
(hereafter B) presents a new opportunity to study this model. We use
the current spin parameters of the two pulsars, together with models
for their spin-down evolution, to place constraints on the age of the
system.  The motivation for this work is twofold. Firstly, a better
constraint on the system age would reduce the uncertainties in
empirical studies of the rate of binary neutron star inspirals --- one
of the key sources for gravitational wave observatories \citep[see,
e.g.,][and references therein]{kkl06}.  Secondly, because the age can
be used to directly compute the post-supernova orbital parameters, we
may be able to better constrain B's progenitor mass and in turn
improve our understanding of the formation of this unique binary
system \citep{ps05,wkf+06,std+06}. A preliminary version of these
results was presented \citet{lbf+05}. In this paper, we consider
models which account for the modification of B's spin by A's
relativistic wind.  Following a brief review of the properties of
J0737--3039 in Section \ref{sec:0737}, we derive age constraints based
on various spin-down models for both pulsars in Section
\ref{sec:spindown} and discuss their implications in Section
\ref{sec:discussion}.

\begin{table}
\caption{\label{tab:models} Summary of the model assumptions 
used in this paper.}
\begin{tabular}{ccccc}
\hline
Model &
Interaction &
$n_{\rm A}$ &
$n_{\rm B}$ &
Torque decay \\
 &
modeled? &
 &
 &
\\
\hline
1 & no  &  0.0---5.0 & 1.4---3.0 & none   \\
2 & no  &    3.0     & 3.0       & B: 10 Myr \\
3 & no  &    3.0     & 3.0       & B: 100 Myr\\
4 & yes &  0.0--5.0  & 1.0       & none \\
5 & yes &  0.0--5.0  & 2.0       & none \\
\hline
\end{tabular}
\end{table}

\section{J0737--3039 and the standard model}
\label{sec:0737}

Applying the binary recycling scenario to the double pulsar system, we
identify A as the first-born neutron star which was spun up (recycled)
by the mass accretion process. B is then the neutron star formed
during the supernova explosion of the secondary.  Using the observed
spin parameters of both pulsars to estimate their surface magnetic
fields $B_{\rm surf} = 3.2 \times 10^{19} \, {\mbox G} \,
\sqrt{P\dot{P}}$, we find $B_{\rm surf,A}=6.3 \times 10^9$ G and
$B_{\rm surf,B}=1.2 \times 10^{12}$ G. That A's field is some three
orders of magnitude lower than that of B is in accord with the
recycling hypothesis --- i.e.~the weaker field of A is a consequence
of the recycling process. We note however, that due to the interaction
of A's wind which penetrates deep into B's magnetosphere (Lyne et
al.~2004), some care should be taken when interpreting the exact value
of B's magnetic field. \citet{absk05} have proposed a model in which
A's wind exerts a propellor torque on B which dominates its spin-down.
In this case, the implied magnetic field strength of B is a factor of
three lower than the above dipole estimate.

After the accretion phase, it is assumed that both neutron stars have
been spinning down due to a steady braking torque; as such they
represent independent clocks measuring the time since accretion
ceased.  A straightforward test of the prediction is to use the
characteristic ages of A and B: $\tau_{\rm A}=P_{\rm A}/(2
\dot{P}_{\rm A})$ and $\tau_{\rm B}=P_{\rm B}/(2 \dot{P}_{\rm B})$.
Lyne et al.~(2004) find $\tau_{\rm A}=2.1 \times 10^8$ yr and
$\tau_{\rm B}=0.5 \times 10^8$ yr.  Possible explanations for this
discrepancy are: (i) the standard evolutionary scenario does not
apply; (ii) as observed in other pulsars \citep[see, e.g.,][]{klh+03}
characteristic ages are not reliable due to their simplifying
assumptions; (iii) both the model and the characteristic ages are
wrong.

Given the circumstantial evidence in favour of the recycling
hypothesis, and the absence of viable alternative models, the simplest
resolution is option (ii). In the rest of this paper, we investigate
the consequences of this case and show that, when the simplifying
assumptions of the characteristic ages are taken into account, the
apparent age differences of the two pulsars can be reconciled.

\section{Modeling the spin evolution and ages of A and B}
\label{sec:spindown}

Our goal is to use the observed parameters and models for the
spin-down of the two neutron stars in J0737--3039 to place constraints
on the system age and pulsar birth parameters. Before describing
specifics, we first outline our general approach. For each pulsar, we
consider a generic spin-down model of the form
\begin{equation}
\label{equ:generic}
\dot{P} = K P^{2-n},
\end{equation}
where $P$ is the spin period, $n$ is the braking index (for spin-down
due to magnetic dipole radiation, $n=3$) and the factor $K$ depends on
the neutron star moment of inertia, braking torque applied to the star
and, for some models of B's spin evolution, the effect of A's
relativistic wind. To distinguish between each pulsar, we add A and B
subscripts where appropriate.  Our basic approach is to apply and
solve the equation under certain model constraints to find the
`spin-down' age of each pulsar, $t_{\rm sd,A}$ and $t_{\rm
sd,B}$. Table \ref{tab:models} summarizes the different models we
investigated.

\begin{figure}
\psfig{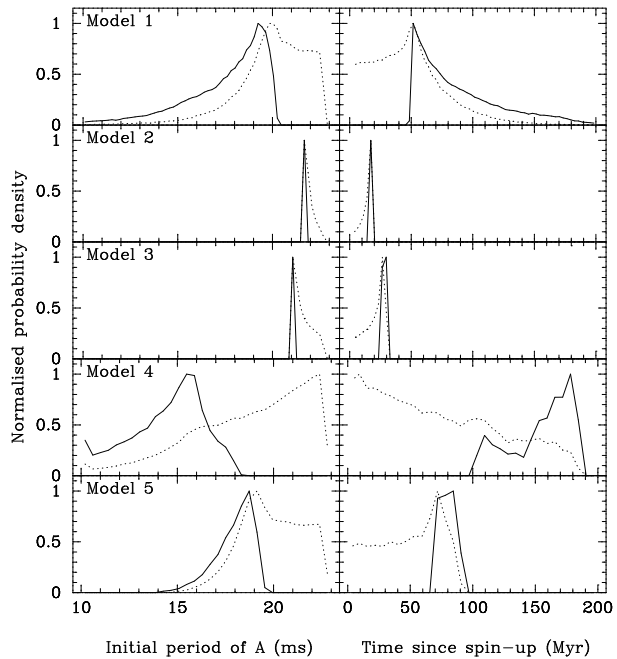}
\caption{\label{fig:distributions}Probability density functions
showing the post-accretion spin period of A (left panels) and the time
since the end of the spin-up phase (right panels). For each model, we
assume a range of initial spin periods for B to be anywhere between
zero and B's current period (dotted curves), and between zero
and 150~ms (solid curves).}
\end{figure}

A key assertion we then make is that the spin-down age for A, $t_{\rm
sd,A}$, refers to the time since spin-up ceased, and is essentially
the same epoch as B began life as a pulsar. We therefore assume
\begin{equation}
\label{equ:assume}
t_{\rm sd,A}=t_{\rm sd,B}.
\end{equation}
Since this condition is only met for certain sets of birth parameters,
we can use it to find the most probable age of the system assuming a
given set of model assumptions.  Our results are summarized in
Fig.~\ref{fig:distributions} and discussed in detail in the following
subsections.

\subsection{Constant braking parameters}

In the simplest case, where $K$ and $n$ are independent of time,
equation (\ref{equ:generic}) can be integrated directly to find
$t_{\rm sd}$. For the case $n = 1$, we find
\begin{equation}
\label{equ:sd1}
t_{\rm sd} = 2 \tau \ln\left(\frac{P}{P_0}\right),
\end{equation}
while, for all other values of $n$, the solution is
\begin{equation}
\label{equ:sd2}
t_{\rm sd} = 
\frac{2 \tau}{(n-1)} \left[1-\left(\frac{P_0}{P}\right)^{n-1}\right].
\end{equation}
Here, $P_0$ is the initial spin period and the
characteristic age $\tau=P/2\dot{P}$.  Since the current period and
period derivative are readily observable through timing observations,
the unknown parameters are $n$ and $P_0$ for each pulsar.

We explore the parameter space using a Monte Carlo simulation to
compute the spin-down age for B assuming an initial spin period and
braking index.  Then, assuming a braking index for A, we compute the
required initial spin period of A and the age of the system by
asserting equation (\ref{equ:assume}).  As a starting point, hereafter
known as model 1, we adopt a braking index for B based on observations
of other normal pulsars \citep[see, e.g.,][for a review]{kh02} which
are consistent with a flat distribution in the range $1.4<n_{\rm
B}<3.0$. Given the completely unknown braking indices for recycled
pulsars in general, for A we took a more conservative approach and
assumed a flat distribution in the range $0<n_{\rm A}<5$.  To show the
effect of B's unknown initial spin period on the results, we performed
all simulations using a flat distribution in the range $0<P_{\rm
0,B}<P_{\rm B}$ (dotted lines) and for $0<P_{\rm 0,B}<150$~ms (solid
lines). The upper bound in the latter case is taken from the range of
initial spin periods inferred from pulsars with multiple age
constraints \citep[see, e.g.,][]{mgb+02,klh+03}.  The resulting
initial spin period distribution for A peaks just below 20 ms and the
age distribution peaks at $\sim 50$ Myr (i.e.~B's characteristic age).

\subsection{Exponential torque decay}

Model 1 assumes no decay of the braking torque.  An alternative
solution to equation (\ref{equ:generic}) can be found for the case
where the braking torque $K$ decays with time.  Assuming, for
simplicity, an exponential decay of the form $K=K_0 \exp(-t/t_{\rm
decay})$, where $K_0$ is a constant and $t_{\rm decay}$ is the $1/e$
decay time, equation (\ref{equ:generic}) integrates to yield the
so-called ``reduced age'', i.e.:
\begin{equation}
\label{equ:sd3}
t_{\rm reduced} = t_{\rm decay} \ln(1 + t_{\rm sd}/t_{\rm decay}).
\end{equation}
Here, $t_{\rm sd}$ is described by either equation (\ref{equ:sd1}) or 
(\ref{equ:sd2})
depending on the assumed value of $n$. 

The cause and even existence of torque decay in isolated non-recycled
neutron stars is uncertain and controversial \citep[see,
e.g.,][]{bwhv92}.  In young neutron stars, it is thought to be due to
either the decay of the magnetic field and/or the alignment of the
spin and magnetic axes with time \citep[see, e.g.,][]{tk01}.  Since
torque decay is not thought to be significant for recycled pulsars
after the accretion phase \citep[see, e.g.,][]{bv91}, 
we consider only the case in
which the torque on B decays. In models 2 and 3 the simulated
distributions shown in Fig.~1 result from the equality $t_{\rm
sd,A}=t_{\rm reduced,B}$ assuming pure magnetic dipole braking
($n_{\rm A}=n_{\rm B}=3$) and a torque decay on B with a timescale of
10 Myr in model 2 and 100 Myr in model 3. In both cases, the time
since spin-up ceased is smaller than for model 1, with the age
distribution peaking at 20--30 Myr.

\subsection{Interaction models}
\label{sec:interact}

So far we have assumed the spin-down of both pulsars to be
independent. In reality, as noted by \citet{lbk+04}, A's rate
of loss of spin-down energy is 3000 times that of B; this, together with
the close proximity of the two pulsars in their orbit, means that 
B's spin-down is significantly affected by A's relativistic
wind.  Direct observational evidence for such an interaction was
presented by \citet{mll+04}.

To model these effects, we follow the results of \citet{lyu04} and
consider two cases. In the first, hereafter model 4, it is assumed
that all the Poynting flux from B is dissipated when it reaches the
interface between A's wind and B's truncated magnetosphere.  Using
equation (9) from \citet{lyu04} for this case, we find that
\begin{equation}
\label{equ:interact1}
\dot{P}_{\rm B} = k_1
               \left(\frac{\dot{E}_{\rm A}}{D^2}\right)^{1/3} P_{\rm B},
\end{equation}
where $k_1$ depends on B's radius and intrinsic magnetic field, $\dot{E}$
is the spin-down energy loss rate of A and $D$ is the separation of
the two pulsars. The basic spin evolution of B in this model implies a
braking index $n_{\rm B}=1$ which is modified by A's spin-down energy
loss. Such a dependence can also be found from the model of
\citet{absk05}.

In the second case put forward by \citet{lyu04}, hereafter model 5,
the interface is partially resistive and large surface currents
produced combine with a poloidal magnetic field of B to produce a
spin-down torque. This process results in a relationship of the form
\begin{equation}
\label{equ:interact2}
\dot{P}_{\rm B} = k_2 \left(\frac{\dot{E}_{\rm A}}{D^2}\right)^{1/2},
\end{equation}
where $k_2$ also depends on B's radius and magnetic field strength.
In this case, the spin evolution of B is independent of its own period
(corresponding to a braking index $n_{\rm B}=2$) and is completely
dominated by A's spin-down energy loss.

\begin{figure}
\psfig{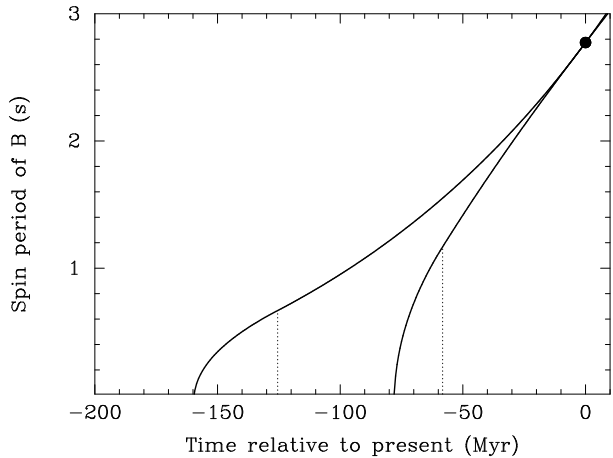}
\caption{\label{fig:tinteract}
The spin history of B assuming two cases (left: model 4; right:
model 5) of magnetospheric interaction from
\citet{lyu04}. B's current spin period and epoch is marked by the
solid point at the top right. The vertical dotted lines in each case
show the point at which A's wind no longer penetrates into B's
magnetosphere (125~Myr for model 4 and 58~Myr for model 5).
}
\end{figure}

To derive age constraints for models 4 and 5, we adopt a slightly
different approach since equations \ref{equ:interact1} and
\ref{equ:interact2} cannot be integrated analytically. Instead, we
solve for the spin period evolution of B numerically assuming the
parameters $k_1$ and $k_2$ to be constant.  Starting with the
currently observed spin parameters, we step back in time and calculate
the variation of $\dot{E}_{\rm A}=4 \pi^2 I_{\rm A} \dot{P}_{\rm A}
P_{\rm A}^{-3}$ using equation (1) to compute $P_{\rm A}$ and
$\dot{P}_{\rm A}$ at each epoch. We assume the canonical neutron star
moment of inertia $I_{\rm A} = 10^{38}$~kg~m$^2$. Simultaneously, we
evaluate $D$ using the results of \citet{pm63} and \citet{pet64}.
During the calculation, we also keep track of whether A's wind
continues to penetrate B's magnetosphere using equations (10) and (12)
from \citet{lyu04}\footnote{Note that there is a missing $c$ in the
numerator of equation (10) of \citet{lyu04} to calculate the magnetic
field strength of B.}. At the point when this condition is no longer
met, we follow B's spin evolution using the standard spin-down formula
given in equation (\ref{equ:generic}).  Example evolution curves are
shown in Fig.~\ref{fig:tinteract} for the case of underlying dipolar
spin-down ($n_{\rm A} = n_{\rm B} = 3$).

The results of our Monte Carlo simulations for models 4 and 5 are
shown alongside the other models in Fig.~\ref{fig:distributions}.
Unlike the other models, the system age constraints depend strongly on
B's unknown birth spin period. Assuming B's period was $<150$~ms at
birth, the age distributions peak sharply at 180 and 80 Myr
respectively for models 4 and 5. For a broader range of initial spin
periods, the system age becomes less well constrained. 
The reason for this can be seen in Fig.~\ref{fig:tinteract} which
shows the large range of system ages possible for a given $P_{\rm
0,B}$.  

\section{Discussion}
\label{sec:discussion}

We have considered a range of spin down models to place age
constraints in the double pulsar system.  A striking result of this
study is the variety of possible system ages. These range from $\lapp
20$~Myr (model 2) to almost 200~Myr (model 4). Models 1--3, which do
not account for the effect of A's wind on B, favour significantly
smaller ages than models 4 and 5 which do account for the interaction.
All models we considered favoured an initial spin period for A that is
close to its currently observed value, with the peak of the
distribution in the range 15--20 ms. This range is consistent with A's
initial spin period predicted by accretion spin-up models \citep[see,
e.g.,][]{acw99}.  Given the current evidence in favour of initial spin
periods for normal pulsars to be $<150$~ms \citep[see,
e.g.,][]{mgb+02,klh+03}, and for interaction between A and B
\citep{mll+04}, we prefer the constraints provided by the solid lines
for models 4 and 5 shown in Fig.~\ref{fig:distributions}.

Since the age of J0737--3039 determines its contribution to the
coalescence rate of neutron star binaries, we briefly revisit the
results of the calculations most recently carried out by \citet{kkl06}
where an age of 230~Myr was assumed.  Taking into account the
additional time to coalescence of 85~Myr, estimates of the total
lifetime of J0737--3039 in this paper range between 105~Myr and
265~Myr. From Table~1 of \citet{kkl06}, we find that the contribution
to the global merger rate made by J0737--3039 would therefore either
increase by 100\% in the youngest case, or drop by 10\% in the oldest
case. In our opinion, the most realistic spin-down models we have
considered are those which take into account the interaction, and
assume that B's initial spin period was negligible compared to its
current value. For model 4, the estimated lifetime would be $\sim
80+85=165$~Myr, i.e.~the merger rate contribution would increase by
40\% over the value found by Kim et al., whereas for model 5, the
contribution does not change significantly. Given the uncertainties in
merger rate estimates, our results do not change the conclusions that
binary neutron star inspirals are unlikely to be detectable by the
current gravitational wave detectors. However, the prospects for
detection by future instruments such as advanced LIGO are excellent.

Our results can be used to constrain the post-supernova orbital
parameters in the double pulsar system. Using the formulae given by
\citet{pm63} and \citet{pet64} we find the mean orbital separation and
eccentricity after the formation of B to be respectively $1.0 \times
10^9$~m and 0.14 for an age of 180~Myr favoured by model 4.  For the
80~Myr solution from model 5, the corresponding numbers are $0.9
\times 10^9$~m and 0.11. These constraints in turn restrict the
allowed ranges of system parameters at the time of the second
supernova \citep{ps05,wkf+06,std+06}.  We have repeated the analysis
of \citet{tds05} who constrain the likely kick velocity at the time
of the second supernova, $V_k$, the tilt angle betweem the pre and
post-supernova orbital planes, $\delta$, and the pre-supernova mass of
B's progenitor star, $m_{2.i}$. These simulations were carried out 
assuming the two different prior distributions of 
the current (unknown) radial
velocity of PSR~J0737$-$3039, as described in detail by \citet{std+06}.

We have imposed our two possible age solutions to the simulations
described by \citet{std+06} by restricting the age ranges considered
to be either 70--90~Myr, or 170--190~Myr as opposed to the range of up
to 100~Myr originally considered by \citet{std+06}.  In general, we
obtain consistent results to \citet{std+06} which indicates that their
work does not critically depend on the system age. The only exception
is the 70--90~Myr simulation assuming the Gaussian radial velocity
distribution, for which the age constraints favour slightly lower
$m_{2,i}$, $V_k$ and $\delta$ values than the unconstrained case.
The new age constraints derived
here are fully consistent with the idea that B's progenitor was a
low-mass star, and that the system received a relatively small
impulsive kick at the time of the second supernova \citep{pdl+05}.

\section*{Acknowledgments}

IHS holds an NSERC UFA and pulsar research at UBC is supported by 
a Discovery Grant. We thank Jonathan Arons and
Bart Willems for useful discussions.
The simulations investigating the companion mass, kick and tilt
angle constraints were
carried out on
a computer cluster funded by a CFI New Opportunities grant to IHS
and M.~Berciu. SET and RJD are supported by the National Science
Foundation under Grant 0506453.

\end{document}